# Low-energy domain wall racetracks with multiferroic topologies

Arundhati Ghosal[1], Alexander Qualls[1], Yousra Nahas[2], Shashank Ojha[3], Peter Meisenheimer[4], Shiyu Zhou[5], Maya Ramesh[6], Sajid Husain[3], Julia Mundy[7], Darrell Schlom[6,8,9], Zhi Yao[10], Sergei Prokhorenko[2], Laurent Bellaiche[2,11], Ramamoorthy Ramesh[3,4,8], Paul Stevenson[12,*], and Lucas Caretta[1,5,*]

[1]*School of Engineering, Brown University, Providence, RI 02912 USA*
[2]*Smart Ferroic Materials Center, Physics Department, Institute for Nano-science and Engineering, University of Arkansas, Fayetteville, Arkansas 72701, USA*
[3]*Department of Materials Science and Nanoengineering, Rice University, Houston, TX 77005 USA*
[4]*Department of Materials Science and Engineering, University of California, Berkeley, CA 94720 USA*
[5]*Department of Physics, Brown University, Providence, RI 02912 USA*
[6]*Department of Materials Science and Engineering, Cornell University, Ithaca, NY 14853 USA*
[7]*Department of Physics, Harvard University, Cambridge, MA 02138 USA*
[8]*Kavli Institute at Cornell for Nanoscale Science, Cornell University, Ithaca, NY, USA*
[9]*Leibniz-Institut für Kristallzüchtung, Berlin, Germany*
[10]*Lawrence Berkeley National Laboratory, Berkeley, CA 94720 USA*
[11]*Department of Materials Science and Engineering, Tel Aviv University, Ramat Aviv, Tel Aviv 6997801, Israel*
[12]*Department of Physics, Northeastern University, Boston, MA 02115 USA*

*Correspondence to: lucas_caretta@brown.edu, p.stevenson@northeastern.edu

**Abstract:**

**Conventional racetrack memories move information by pushing magnetic domain walls or other spin textures with spin-polarized currents[1,2], but the accompanying Joule heating inflates their energy budget and can hamper scaling. Here we present a voltage-controlled, magnetoelectric racetrack in which transverse electric fields translate coupled ferroelectric–antiferromagnetic walls along $BiFeO_3$ nanostrips at room temperature. Because no charge traverses the track, the switching dissipates orders of magnitude less energy than the most efficient spin-torque devices with more favourable scaling, making the scheme significantly more attractive at the nanoscale. We further uncover noncollinear topological magnetoelectric textures that emerge at domain walls in $BiFeO_3$, where the nature of these topologies influences their stability upon translation. Among these are polar bi-merons and polar vertices magnetoelectrically coupled with magnetic cycloid disclinations and previously unobserved, topological magnetic cycloid twist topologies. We observe domain wall velocities of at least kilometres per second—matching or surpassing the fastest ferrimagnetic and antiferromagnetic racetracks[3,4] and approaching the acoustic-phonon limit of $BiFeO_3$—while preserving these topologies over tens of micrometres. The resulting high velocity, low-energy racetrack delivers nanosecond access times without the thermal overhead of current-driven schemes, charting a path toward dense, ultralow-power racetrack devices which rely on spin texture translation.**

**Introduction:**

Topological and chiral solitons—such as skyrmions, Bloch points, merons, and domain walls (DWs)—have emerged as promising candidates as information carriers in next-generation memory, logic, and quantum information science applications[1,2,5–11]. Efficient translation and manipulation of these spin textures at high speeds is critical for realizing racetrack-based devices. In this context, DWs and skyrmions in compensated magnetic systems, including ferrimagnets and antiferromagnets (AFMs), have garnered increasing attention due to their intrinsic advantages, such as fast spin texture dynamics and immunity to stray magnetic fields[4,12–22]. Experimental demonstrations of spin texture translation in compensated ferrimagnets and synthetic AFMs have underscored these benefits[4,12,13,16–20]. Most demonstrations of spin texture motion, however, have relied on current-induced spin torques in metallic systems, which are energetically costly due to significant Joule heating and the need for high critical current densities[19]. Although magnon-mediated translation has been proposed as a lossless and energy-efficient alternative[23–25], the generation of magnons typically still requires inefficient current-based excitation—either through coherent RF striplines or incoherent spin torques.

In parallel, noncollinear textures in ferroelectric materials—so-called *polar textures*—have gained traction in single-layer, heterostructure, and superlattice geometries[26–31], where their nucleation and annihilation have been demonstrated using low-power electric fields. These polar texture topologies exhibit rich physical phenomena, including emergent chirality[26,27,32,33] and local negative capacitance[34,35], offering an alternative platform for topological solitons with new benefits. Dynamical control and long-range translation of polar textures across device-relevant length scales, however, remain largely elusive and understudied. Combining the often mutually exclusive advantages of antiferromagnetic and ferroelectric noncollinear textures—namely, ultrafast dynamics, insensitivity to stray fields, and low-power electric control—represents a central challenge in the development of next-generation, energy-efficient racetrack-like devices.

In this context, magnetoelectric materials, such as $BiFeO_3$ (BFO), which exhibit strong coupling between antiferromagnetic and ferroelectric orders at room temperature, provide a compelling opportunity. These systems enable electrical control over antiferromagnetic order and offer a potential route to nucleate and manipulate nonlinear spin textures without relying on current-induced effects. Recent studies have shown that applying radial electric fields to multiferroic BFO thin films can nucleate or annihilate antiferromagnetic singularities[36]. Nonetheless, while these early demonstrations highlight the potential of electric-field control, the controlled translation of magnetoelectric topologies remains largely unexplored. Unlocking this capability is an essential step towards a new paradigm of electrically reconfigurable and energy-efficient topological devices that leverage the dual ferroic nature of magnetoelectric materials.

Here, we combine the advantages of AFM spin texture dynamics with low-power ferroelectric manipulation in room-temperature magnetoelectric BFO to create novel racetrack devices driven by transverse electric fields. We utilize the coexistent order parameters in BFO to stabilize topological noncolinear magnetoelectric textures and drive them into motion with electric fields. By correlating piezoresponse force microscopy (PFM) and scanning nitrogen vacancy (NV) magnetometry, we discover that multiferroic DW boundaries in BFO can give rise to novel, real-space magnetoelectric topological textures, which combine polar bi-meron and polar vertex ferroelectric topologies with distinct magnetic topologies, including previously unobserved AFM cycloid knot topologies arising from the Klein bottle geometry[37] of the underlying order parameter space. The emergence of magnetoelectric topologies is explained through atomistic effective Hamiltonian simulations, which reveal a complex interaction of individual spin cycloids at the magnetoelectric domain boundary.

We demonstrate robust electric field driven motion of magnetoelectric DWs and topologies in prototypical ratchet and racetrack devices, where the magnetoelectric topologies are driven at conservative estimate of 1000 m/s – close to the acoustic phonon group velocity of BFO[38] – and energies that are 1-2 orders of magnitude lower than spin-orbit torque or spin-transfer torque racetracks at similar or even lower velocities than reported here. Finally, by combining experiments and simulations, we characterize the stability of the magnetoelectric topologies and their robustness to translation across tens of micrometers. This work offers fundamental insights into the formation, stability, and electric field-driven dynamics of magnetoelectric topologies, while providing a materials and device design strategy for low-power racetrack computational devices utilizing room-temperature magnetoelectric coupling.

**Engineering topology at magnetoelectric domain walls**

Bismuth ferrite (BiFeO$_3$ or BFO) is widely recognized as a leading candidate for low-power computation due to its unique room-temperature magnetoelectric properties. In bulk, BFO crystallizes in the rhombohedral *R3c* space group, with a spontaneous ferroelectric polarization, **P**, oriented along <111> directions. It also hosts a long-wavelength, incommensurate spin cycloid with a period of approximately 64 nm (see Fig 1a). This type-I spin cycloid propagates along $\langle 1\bar{1}0\rangle_{PC}$ directions that are perpendicular to **P**[39–43] and arises from a Dzyaloshinskii–Moriya interaction (DMI) stemming from ferroelectric structural distortion, a so-called "polar" DM[44–46]. Similarly, a type-II spin cycloid, which propagates along $\langle 11\bar{2}\rangle_{PC}$ directions has also been observed in BFO under certain epitaxial boundary conditions[43]. In thin-film geometries, epitaxial strain and electrostatic boundary conditions break the degeneracy of domain variants, leading to a diverse array of ferroic domain structures. The most common being, in (001)-oriented films, 71° and 109° DWs (illustrated in Fig. 1a), which differ primarily by the orientation of the out-of-plane component of **P** and correspond to specific ferroelectric/ferroelastic switching pathways[47,48]. [43]

The propagation direction of the spin cycloid is closely tied to the orientation of **P** in each domain, underscoring the strong coupling between electric and magnetic order in BFO[42,43,48,49]. Figures 1b and 1c show these polar domain configurations: in-plane and out-of-plane vector-based piezoresponse force microscopy (PFM) images reveal characteristic striped ferroelectric domains separated by 71° (Fig. 1b) and 109° (Fig. 1c) DWs that emerge when ~100 nm-thick (001)-oriented BFO films are epitaxially grown on orthorhombic DyScO₃ (DSO) substrates (Methods). See Extended Data Figs. 1 and 2 and Supplementary Methods for further details. Complementary scanning nitrogen-vacancy (NV) magnetometry images (Fig. 1d and e) capture the local stray magnetic fields emanating from the spin density wave superimposed with the AFM spin cycloid in BFO. These show the now characteristic zig-zag magnetic pattern perpendicular to the projection of **P** [42,43,48,49], specific to different domain configurations[50]. Insets in Figs. 1d and e correlate the local in-plane PFM signal from Fig 1b and c with the cycloid phase, confirming a one-to-one correspondence between polar and cycloid domains[42,43,48,49]. Of particular interest is the seamless rotation of the cycloid propagation vector, **k**, across ferroelectric DWs in both samples.

While this versatility provides a rich design space for engineering nontrivial ferroic textures at domain boundaries, most studies of coupled polarization and magnetization in BFO have focused on interfaces between so-called $n = 2$ domains[42,43,48,49], such as those shown in Figs. 1b-e. However, Figs. 1f and g show that by engineering interfaces involving $n > 2$ domain intersections, we can experimentally realize nontrivial and complex magnetoelectric orderings. We identify two distinct classes of such features – polar vertex-like (Fig. 1f) and polar bi-meron-like (Fig. 1g) structures, named to reflect similarities with previously simulated ferroelectric polarization textures[51]. Figure 1f presents an in-plane PFM image of the same ~100 nm-thick, $(001)_{PC}$-oriented BFO film grown on DSO $(110)_O$ shown in Fig. 1b, highlighting a distinct region of the sample where $n = 4$ separate 71° ferroelectric DWs converge to form a topological vertex (green circle). Here PC and O subscripts denote pseudocubic and orthorhombic notation, respectively. This polar vertex possesses a net winding number $W_{FE} = 1$, with local **P** directions indicated by black and white arrows. The inset schematically illustrates the winding structure of the vertex. Surprisingly, NV magnetometry reveals a previously unobserved AFM knot topology in the magnetic ordering, comprised of a vortex of the cycloid propagation vector, **k**, and its phase $\phi$, which we further justify later and whose 3D spin texture is shown in Fig. 1a for clarity. Unlike conventional cycloids found at $n = 2$ DWs, where the propagation direction smoothly rotates across adjacent polar domains, $n = 4$ vertices exhibit a frustrated and nonuniform cycloid configuration. Tailoring the domain convergence geometry likewise generates vertices with the opposite winding number ($W_{FE} = -1$; Extended Data Fig 3). In addition to fourfold vertices, nontrivial structures also arise from the intersection of $n = 3$ unique ferroelectric domains, as shown in Fig. 1g (blue circle). Here, domain convergence results in the formation of polar meron pairs ($W_{FE} = ½ + ½$), highlighted in blue and schematically shown in the inset. Scanning NV magnetometry (Fig. 1g) shows that these bi-meronic textures exhibit their own distinct AFM spin topology, described as disclinations discussed in detail later. Additional examples of both vertex and bi-meron polar topologies, as well as their associated knot and disclination spin topologies, are provided in Extended Data Fig 4.

These $n = 3$ and $n = 4$ magnetoelectric topologies naturally form at the boundaries between regions where the net in-plane polarization in BFO changes by 180°, indicated by red and purple arrows in Figs. 1e and 1g and examined in greater detail in Extended Data Fig 4. DWs with antiparallel in-plane components can be generated and manipulated in BFO either by applying lateral electric fields (as demonstrated here and detailed later) or through epitaxial boundary conditions. Particularly, the ability to extrinsically manipulate and translate these emergent magnetoelectric solitons with electric fields offers a powerful and energy-efficient pathway to combine the advantages of ferroelectric and AFM orders in racetrack-like devices.

**Magnetoelectric translation of magnetic domain walls and noncollinear topologies**

In-plane 180° magnetoelectric DWs in BFO can be readily translated through the application of a lateral electric field oriented either parallel or antiparallel to the polarization vector of one of the constituent domains. Leveraging this mechanism—along with the observation that the emergent $n = 3$ and $n = 4$ magnetoelectric topologies are inherently anchored to these mobile DWs—we demonstrate the ability to translate multiferroic topologies using voltage pulses. To showcase this capability, we fabricate prototype ratchet-style DW racetracks, a device geometry previously proposed for shift-register-like memory and neuromorphic computing applications[52–54]. In our implementation, lateral electrodes are patterned orthogonal to the net in-plane polarization of the 71° BFO domains and aligned parallel to the normal of the 180° DW, as illustrated in Fig. 2a. The asymmetric ratchet geometry enables deterministic control over DW nucleation and translation: for a given applied voltage, the electric field concentrates more strongly at the narrow, pinched side of the device (left side in Fig. 2a) than at the wider side (right side in Fig. 2a). This field gradient ensures that DW nucleation initiates at the narrow end, while subsequent in-plane electric field pulses drive the 180° ferroelectric DW along the track toward the wider side (Fig. 2a). Through this approach, we achieve controlled, voltage-driven motion of both the DW and the associated complex magnetoelectric textures.

Figure 2b presents overlaid in-plane PFM images acquired on a fabricated asymmetric DW racetrack, implemented on the same BFO sample used to demonstrate the nontrivial magnetoelectric topologies described above. The lateral electrodes, shown schematically in grey, define the racetrack boundaries. Each overlaid image corresponds to a sequential voltage pulse, driving a single nucleated magnetoelectric DW along the racetrack. This time-lapse series confirms our hypothesis that the ratchet-style device facilitates deterministic nucleation and propagation of a single 180° DW, enabling unambiguous tracking of complex magnetoelectric textures under consecutive electric field pulses. Figures 2c–f display higher-resolution lateral PFM and scanning NV magnetometry measurements focused on a region of interest within the DW (highlighted by the white box in Fig. 2b). Strikingly, we observe that vertex-like structures ($n = 4$) are robustly translated over the total length of the ratchet device (~20 $\mu m$) with minimal distortion. The swirling magnetic knot texture, including its sense of rotation, also remains preserved throughout the translation process. —even as the DW encounters material defects (see Extended Data Figs. 5-8). Figure 3a and Extended Data Figs. 5-8 track the evolution of the interfacial topological objects during successive voltage pulses in ratchet devices. The $n = 4$ vertex and corresponding AFM knot translates coherently over distances exceeding tens of micrometers, whereas the $n = 3$ bi-meron and disclination pairs are frequently disrupted by local perturbations of the ferroelectric background. These features tend to annihilate, merge, or newly nucleate during motion, making it difficult to track individual bi-meron pairs as the DW propagates. We attribute the superior robustness of the $n = 4$ interface, in part, due to differences in the energy costs of 71° vs 109° switching events; the $n = 4$ interface requires only 71° switching steps to propagate while $n = 3$ interfaces require both 71° and 109° steps (Extended Data Fig. 9, Supplementary Discussion).

This distinction highlights the enhanced robustness and reliability of stabilized $W_{FE} = 1$ vertex structures and their corresponding magnetic topologies. Similar behavior is reproducibly observed in a second DW racetrack device, shown in Extended Data Figs. 5 and 7. We note that in the final DW state shown in Fig. 2f, the magnetic cycloid swirl appears slightly distorted. We attribute this distortion to the DW translating beyond the lithographically defined active region of the device, where the applied electric field becomes poorly defined, potentially leading to more stochastic domain formation and propagation.

**Origins of nanoscale topological magnetoelectric textures**

To explain the emergence and stability of the magnetoelectric textures observed at $n > 2$ domain intersections, we analyze the AFM order of BFO using a combined framework of homotopy theory and atomistic effective Hamiltonian simulations (see Methods). As detailed in the Supplementary Discussion and Extended Data Figs.10-14, the range of cycloidal spin states associated with a given polarization state in BFO is homotopic to the polygonal representation of a Klein bottle (KB). As a result, this formalism conveniently enables a rigorous classification of the topological AFM defects that nucleate where multiple domains with varying **P** meet. For $n = 3$ junctions, homotopy theory predicts appearance of elementary defects – the +1/2 ("i"-shaped) and -1/2 ("y"-shaped) disclinations shown via Néel vector maps in Fig. 3b and c, respectively. To minimize the distortion of the spin pattern and accommodate topologically trivial cycloid states on either side of the field induced domain boundary such disclinations appear in tightly bound +1/2 and -1/2 pairs. This is confirmed by our effective Hamiltonian simulations which show that the disclinations correspond to minimal energy spin states at polar merons. Experimentally, stray-field maps obtained with scanning NV magnetometry are consistent with the stray field calculated from the AFM cycloidal disclination (Fig 3d,e), shown as a sequence of "i" $(+1/2)$ and "y" $(-1/2)$ patterns along the $[001]_O$ DW in BFO (Fig. 3f), and similar to those observed elsewhere[55,56] (see Extended Data Fig. 12 for additional $\pm 1/2$ disclinatons).

By contrast, $n = 4$ junctions give rise to a markedly richer spin structure. The spiral-like topologies observed by scanning NV magnetometry can be theoretically reproduced when dissimilar cycloid classes– type-I, **k** ∥ [110] and type-II, **k** ∥ [112]–adjoin across the 180° domain boundary in BFO. Under this condition, the energy-minimized solution is a previously undiscovered, twisted defect that can be represented as a non-trivial knot on the KB manifold[37]. Such knot is uniquely formed by a simultaneous vortex of the cycloid propagation vector **k** and a vortex of the cycloidal phase $\phi$ (Supplementary Discussion). This topological spin state shown in Fig. 3g produces a stray field profile (Fig. 3h) consistent with that observed from spiral-like topologies found at $n = 4$ domain intersections (Fig. 3i). The identified topological twisted defects – previously only confined to theoretical predictions in treatments of liquid crystals[57] – are observed here for the first time. The three-dimensional reconstruction of the spin field (Fig. 1a) reveals a compact vortex of the Néel vector threaded by a periodic cycloidal texture. Both of these ingredients are essential to realizing this unique topology: a vortex of the AFM order parameter and a periodic modulation of that order—the cycloid itself. Significantly, because the cycloid modulation exists only in the AFM sublattice and not the polar order, the magnetoelectric object carries distinct topological structures in magnetic and polar orders that are rigidly, magnetoelectrically mapped onto one another. The result is a single nanoscale object that stores two distinct, but inseparable, topological states—establishing a dual-encoded information carrier unprecedented in multiferroic materials.

High-resolution NV magnetometry in Extended Data Fig. 15 experimentally confirms that across the switched 180° elastic boundary, domains host 120°-rotated (type II) or orthogonal (type I) cycloids in the as-grown and electrically switched states, respectively, in full agreement with the simulations (see Supplementary Discussion). We propose that these unique topologies arise due to magnetic frustration resulting from a phase mismatch of **k** and $\phi$ on both sides of the DW from the intersection of differing cycloid types. We note here the correspondence between these results and early description of skyrmion lattices, where spatially localized, non-trivial topological textures can be realized through the superposition of various plane-wave orderings[58–62].

**Energy Efficient Information Transport**

To benchmark the efficiency of voltage-driven topological magnetoelectric (ME) racetracks against state-of-the-art spin-torque (ST) devices, we quantify the primary energy dissipated during DW translation in each architecture. Because ferroelectric order in BFO dictates the magnetic state, the energy required to displace a coupled ferroelectric–AFM wall reduces to $U_{ME} = 4 P_r E_c t_{ME} D \cdot l$, where $P_r$ is the remnant polarization of the ferroelectric, $E_c$ is the ferroelectric coercive field, $t_{ME}$ is the thickness of the ferroelectric, $D$ is the electrode spacing, and $l$ is the length of DW racetrack (Supplementary Discussion, Extended Data Fig. 16). $U_{ME}$ scales linearly with device volume, so shrinking any track dimension proportionally lowers the switching energy. By contrast, the dominant energy loss in ST tracks is Joule heating in the spin-current layer[19], which can be expressed as $U_{\text{ST}} = I^2 Rt = \frac{J_{\text{crit}}^2 l^2 \rho A}{v_{\text{DW}}}$, where $J_{\text{crit}}$ is the critical current density, ρ is the resistivity of the spin current layer, $A$ is the cross sectional area of the wire, and $v_{\text{DW}}$ is the DW velocity. Even under the most optimistic assumptions—operation at $J_{\text{crit}}$ and neglecting all parasitic dissipations, $U_{\text{ST}}$ grows quadratically with $l$ because both resistance and drive time scale with length, exacerbating the heat dissipation in ST racetracks. Extended ST devices capable of storing multiple bits per cell are, therefore, penalized far more severely than their ME counterparts (Supplementary Discussion). Figure 4a illustrates these contrasting scaling behaviors. Solid lines trace the projected volume dependence of $U_{\text{ME}}$; dashed lines show $U_{ST}$ for idealized ST tracks of various fixed racetrack lengths using the best material parameters reported to date, illustrating the detrimental energy effects of extended ST racetracks. Energy budgets extracted from several seminal ST racetrack studies on spin texture motion— spanning diverse materials, geometries- and torque mechanisms—are superimposed for comparison[1,9,14,15,18,19,63–72]. All ST data assume operation at the published $J_{\text{crit}}$, a dwell time set by the fastest reported $v_{\text{DW}}$, and nominal values of $\rho$ for the spin-current-generating material. Further, fabricated devices frequently fall orders of magnitude above the idealized ST trend, particularly at submicron lengths. By contrast, our proof-of-concept ME ratchet (star, Fig. 4a) consumes less energy than nearly all reported ST racetracks and is projected to be almost two orders of magnitude more efficient than ST devices of comparable volume, based on the measured $P_r$ and geometry. Moreover, it is more energy efficient than some ST devices that are two orders of magnitude smaller in size. Although the prototype studied here employs lateral electrodes—inherently more difficult to shrink below the micron scale—the underlying ME concept and energy scaling are agnostic to field geometry. Another significant advantage of the ME structure is that the driving stimulus (electric field) is applied perpendicular to the direction of DW motion, whereas in ST racetracks the stimulus (current) flows along the same axis as the wall. This orthogonality means that the ME architecture can be implemented with an out-of-plane gate stack in which a vertical field drives horizontal wall motion, allowing straightforward lithographic scaling to sub-100 nm dimensions without sacrificing the efficient electric field-driven- mechanism.

However, energy efficiency is only meaningful if the racetrack also meets the DW velocity targets set by ST technology of approximately ≥ 1000 m/s[73]. Since only one DW traverses the device in the ratchet configuration, we can estimate the DW velocity using switching dynamics experiments. We characterize switching dynamics using well-established positive up–negative down- (PUND) measurements (Methods) on ME tracks to extract switching current transients (Fig. 4b, see Methods). The voltage pulse sequence drives real time polarization transients (Fig. 4c) in which DWs move at least 2 μm within the ~2 second saturation time. Because the measurement bandwidth is limited by the RC constant of the device, these data place a conservative lower bound of ∼1000 m/s on the ME DW velocity – on par with the fastest ST devices and approaching the acoustic phonon velocity of BFO (~5000 m s$^{-1}$)[38] . Together, these results establish magnetoelectric racetracks as a compelling low-energy, high-speed alternative to current-driven ST systems.

**Outlook**

The demonstration of voltage-driven magnetoelectric racetracks propelling coupled ferroelectric–antiferromagnetic walls at kilometer-per-second speeds while dissipating order-of-magnitude less energy elevates DW technologies as low-power contenders for next-generation computing. Crucially, the same transverse electric field that delivers this energy–delay advantage also transports distinct topological magnetoelectric motifs—polar four-fold vertex cores and AFM cycloidal knots formed by the convergence of type-I and type-II spin cycloids in BFO. These self-protected textures propagate intact over tens of micrometers, effectively turning each DW into a topologically encoded data packet that can carry multilevel information and act as a reconfigurable element for magnonic or neuromorphic circuits. By uniting capacitive efficiency, magnetic nonvolatility, and intrinsic topology, magnetoelectric racetracks lay a foundation for dense shift-register memories, in-memory logic, and adaptive spin-wave architectures, pushing information technologies toward a regime where speed, endurance and power consumption are no longer at odds.

**Methods:**

*Epitaxial Synthesis of BiFeO₃ thin films:* Thin films of bismuth ferrite (BiFeO$_3$, BFO) of thickness 100 nm and 150 nm were epitaxially deposited on [110]$_O$-oriented dysprosium scandate (DyScO$_3$, DSO) substrates by pulsed laser deposition and molecular-beam epitaxy. The film that was used to study 71° DWs was grown by pulsed laser deposition using a KrF excimer laser (wavelength 248 nm, COMPex-Pro, Coherent). DSO is reasonably well lattice matched with BFO (~0.3%) to ensure epitaxial growth. The substrate was prepared by sonication in acetone followed by isopropyl alcohol and drying with filtered nitrogen. Substrates were attached to the heater with silver paint for good thermal contact. BFO was deposited at a temperature of 710°C under a dynamic oxygen pressure of 140 mTorr and the laser fluence of 1.8 Jcm$^{-2}$ with a 15 Hz laser pulse repetition rate. Post deposition, the films were cooled down to room temperature at 30°C/min at a static O$_2$ at atmospheric pressure.

The film with 109° domains was grown by reactive MBE in a Veeco GEN10 system using a mixture of 80% ozone (distilled) and 20% oxygen. Elemental sources of Bi and Fe were used at fluxes of $1.5 \times 10^{14}$ and $2 \times 10^{13}$ atoms/cm$^2$s, respectively, corresponding to a relative flux ratio of 8:1. All films were grown at a substrate temperature of 675°C as measured by an optical pyrometer operating at a wavelength of 980 nm and a chamber background pressure of $5 \times 10^{-6}$ Torr.

*Electrode Patterning and Deposition:* Device fabrication was done using cleanroom lithography techniques. The mask file was made in autoCAD and written using a Heidelberg MLA 150 Maskless Aligner. An AJA Sputter System (Orion Chamber) was used to deposit the devices with a background base pressure of $\sim 8 \times 10^{-9}$ Torr. The device electrodes are a bilayer of Ta(10 nm)/Pt(70 nm) and were grown at room temperature under a 35 sccm Argon flow and Argon pressure of 3 mTorr. Deposition rates for each element were calibrated using X-ray reflectivity measurements. After electrode deposition, the film was soaked in acetone several hours to initiate liftoff. Subsequently, it was sprayed with acetone using a squirt bottle to agitate any leftover resist. This was followed by an isopropyl alcohol wash.

*Laboratory-based X-ray diffraction:* To verify epitaxial growth, reciprocal space map (RSM) scans were performed on a Panalytical diffractometer with a Cu K-alpha source. RSMs were taken around the (332) DSO diffraction peak along the $[001]_O$ axis to access the $(103)_{PC}$ BFO peaks.

*Piezoresponse Force Microscopy:* The ferroelectric domains in BFO were imaged by piezoresponse force microscopy (PFM) using Asylum Jupiter XR and MFP-3D Origin atomic force microscopes. Out-of-plane components of the polarization were imaged in DART mode and in-plane components were imaged in lateral mode. BFO films exhibit two in-plane polarization variants in the virgin state with a net polarization in the $[001]_O$ direction. The in-plane BFO domain orientations are distinguished by vectorized PFM. The sample was rotated and imaged at 0°, 45°, 90° and 315°, where 0° is the $[1\bar{1}0]_O$ tip scan direction, and measured in an off-resonance condition[74–77] to ensure contrast consistency (see Supplementary Methods).

*Scanning Nitrogen Vacancy Magnetometry:* The antiferromagnetic order of BFO was probed at room temperature using a Qnami ProteusQ, a commercial scanning nitrogen vacancy magnetometer (SNVM), and Quanti-lever MX+ diamond parabolic scanning probe tips with single NV centers. The scanning NV microscope combines a confocal optical microscope with a tuning-fork based atomic force microscope. The NV center, a lattice defect in diamond, is optically excited and the electron Zeeman interaction obtained is sensitive to magnetic fields. Thus, by optically detecting and monitoring the Zeeman shift of the electron sublevels, local magnetic fields can be read.

The devices on the surface of the BFO/DSO films were oriented with the tip scan direction along the DSO $[001]_O$ direction. SNVM data was correlated to PFM data by referencing the topographic imaging modes in both techniques to defects or devices edges. Magnetic textures in BFO were imaged via optically detected magnetic resonance (ODMR) based imaging. The response of the NV center to various microwave field frequencies (typically 12-14 points over a range of 2.90-2.97 GHz were used to minimize acquisition time) was monitored to determine the local magnetic field via the Zeeman shift of the NV center resonance.

*Electrical Ferroelectric Measurements*: To propagate the magnetoelectric DW in our sample, we apply 1 ms voltage pulses using conducting probes and a Radiant Precision Multiferroic Tester. We first apply a 60 V pulse so that the domain nucleates in the 2 μm nucleation zone on one end of the device. The voltage is incrementally increased, and the DW propagates into the ratchet zone of the device until the electric field in the ratchet is less than the critical field for DW motion.

For time-resolved measurement of DW dynamics, we use radio-frequency (RF) voltage probes and a custom-built high-voltage pulse generation circuit that outputs 50 V pulses with 700 ps risetimes. We use these fast pulses to perform positive up–negative down- (PUND) measurements on simple in-plane parallel plate capacitors with 2 μm separation between electrodes. Voltage response measured using a Lecroy SDA 6020 Serial Data Analyzer across the instrument's internal 50 Ω termination to ground. Extracted polarization transients show ferroelectric switching saturates after ~2 ns. We expect these transport measurements to be bandwidth-limited by our voltage probes and device RC constant. We can confidently put a lower bound on the DW velocity at ≥1000 m/s, assuming the switched ferroelectric domains nucleate instantaneously and travel the 2 μm distance between electrodes in the 2 ns switching time.

*First-Principle-Based Calculations and Simulations*: The calculations of stability of various topological spin structures are performed using the first-principle-based effective Hamiltonian model including both structural and magnetic degrees of freedom. The potential energy surface obtained from first-principles calculations is parametrized in terms of local modes $u_i$ (polar displacements) proportional to electric dipole moment of the conventional unit cell $i$, oxygen octahedra tilts $\omega_i$ and displacement vectors $v_i$ characterizing the inhomogeneous strain tensor $\eta_i$. Finally, the structural part of the Hamiltonian also includes the six independent components of the homogeneous strain tensor $\eta_{H_i}$. In terms of magnetic degrees of freedom, the classical magnetic moment vectors $m_i$ are mapped on the $Fe^{3+}$ lattice sites. The coupling between the lattice and magnetic degrees include the DMI interaction induced by oxygen octahedra rotations as well as the spin current term coupling $m_i$ to the polar vectors $u_i$. The former contribution is responsible for the weak magnetization due to the spin canting while the latter term gives rise to cycloidal modulation in the vicinity of the R point ($\frac{1}{2},\frac{1}{2},\frac{1}{2}$) of the pseudo-cubic Brillouin zone. These and other terms included in the effective Hamiltonian are described in Ref.[44]

Using the effective Hamiltonian model, we first perform 40,000 sweep Monte Carlo relaxation of the initial 71° DW structure using a 48x48x12 supercell. The electric-field-induced switching is modeled by reversing the *x* ([100]$_{PC}$) Cartesian components of the polar displacements within half of the supercell, i.e. for sites with *y* ([010]$_{PC}$) coordinates for which *y<24*. The artificially obtained structure is then relaxed for an additional 40,000 sweeps at 300K. The AFM knot structure is obtained by imposing a mismatching type-I and type-II cycloids on different sides of the created ferroelectric domain wall as initial conditions.

**Author Contributions:**

L.C. and P.S. led the study and guided the team.; L.C. and P.M conceived the project; L.C. and P.S. planned the experiments; A.G. performed all PFM measurements and all NV magnetometry with the help of S.O., under the supervision of R.R., P.S., J.M., and L.C.; PLD sample synthesis was optimized by S.H. and P.M., and oxide MBE sample synthesis was optimized by M.R. under the supervision of D.S.; A.Q. performed the energy modeling study and the ferroelectric dynamics measurements with guidance from Z.Y.; Atomistic effective Hamiltonian computations and topology theory was performed by Y.N. and S.P.; L.C., P.S., A.G., S.P. wrote the manuscript; All authors contributed to the discussion of the data in the manuscript and the supplementary materials.

**Acknowledgements:**

Devices were fabricated using equipment in the Brown University Instrumentation for Molecular and Nanoscale Innovation facility. **Funding**: Scanning Nitrogen Vacancy Magnetometry measurements were supported by Air Force Office of Scientific Research under FA9550-24-1-0169. Oxide molecular beam epitaxy synthesis of BiFeO$_3$ was supported by the NSF PARADIM (DMR-2039380) and NSF EPSCOR RII Track-4 Research Fellows Program under OIA-2327352. Theory and simulations effort were supported under Vannevar Bush Faculty Fellowship (VBFF) Grant No. N00014-20-1-2834 from the Department of Defense, Grant No. MURI ETHOS W911NF-21-2-0162 from Army Research Office (ARO), an ARA Impact Grant 3.0, and the MonArk NSF Quantum Foundry supported by the National Science Foundation Q-AMASE-i Program under NSF Award No. DMR-1906383. Oxide molecular beam epitaxy synthesis of $BiFeO_3$ was also supported by Grant No. MURI ETHOS W911NF-21-2-0162 from Army Research Office (ARO). S.H. and R.R. acknowledge support from Rice University, and sustained support of the U.S. Department of Energy, Office of Basic Energy Sciences, Materials Sciences and Engineering Division under Contract No. DE-AC02-05-CH11231 (Codesign of Ultra-Low-Voltage Beyond CMOS Microelectronics) for the development of materials for low-power microelectronics. S.H. and R.R. also acknowledge that this research was sponsored by the Army Research Laboratory and was accomplished under Cooperative Agreement Number W911NF-24-2-0100. The views and conclusions contained in this document are those of the authors and should not be interpreted as representing the official policies, either expressed or implied, of the Army Research Laboratory or the U.S. Government. The U.S. Government is authorized to reproduce and distribute reprints for Government purposes notwithstanding any copyright notation herein. R.R. acknowledges the ARO-CHARM program as well as the NSF-FUSE program.

**Competing interests:** The authors declare no competing interests.

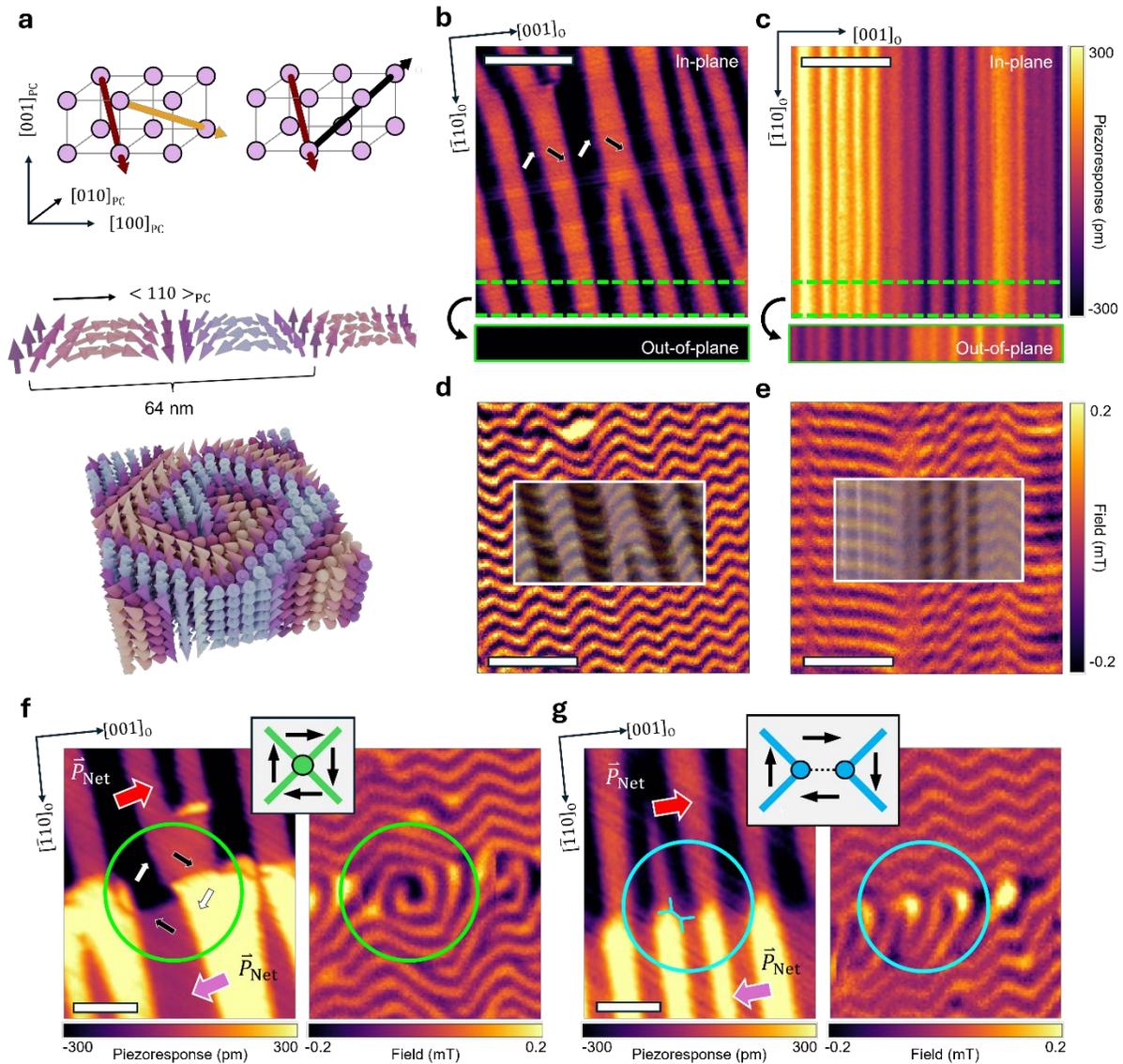

**Figure 1 | Magnetoelectric textures at domain boundaries. a)** Schematic of 71° and 109° BFO DWs, Néel vector map of the BFO cycloid, and 3D rendering of the knotted spin texture. Lateral piezoresponse force microscopy (LPFM) image of a ~100 nm BFO film with **b)** 71° DWs and **c)** 109° DWs grown on a (110)$_O$-oriented DSO substrates. Out-of-plane PFM in the dashed-green areas of the LPFM in (b) and (c) are shown in solid green boxes. Magnetic stray field maps of the **d)** 71° DW and **e)** 109° DW in (b) and (c), respectively, as imaged by scanning NV magnetometry. The scalebars in (b)-(e) are 500 nm. **f)** LPFM and Scanning NV magnetometry of a correlated polar vertex and chiral knot topology with winding number $W_{FE} = +1$ formed when $n = 4$ BFO domains intersect. The inset shows a schematic of the polar vertex. **g)** LPFM and Scanning NV magnetometry of a correlated polar bi-meron and magnetic disclination topology with winding number $W_{FE} = \frac{1}{2} + \frac{1}{2}$ formed when $n = 3$ BFO domains intersect. The inset shows a schematic of the polar bi-meron. Red and pink arrows designate the net polarization direction $\vec{P}_{Net}$ in different regions of the film. Green and blue circles identify the location of each magnetoelectric topology. The scalebar in (f) and (g) is 250 nm. pm, picometers; mT, milliTesla; PC, pseudocubic.

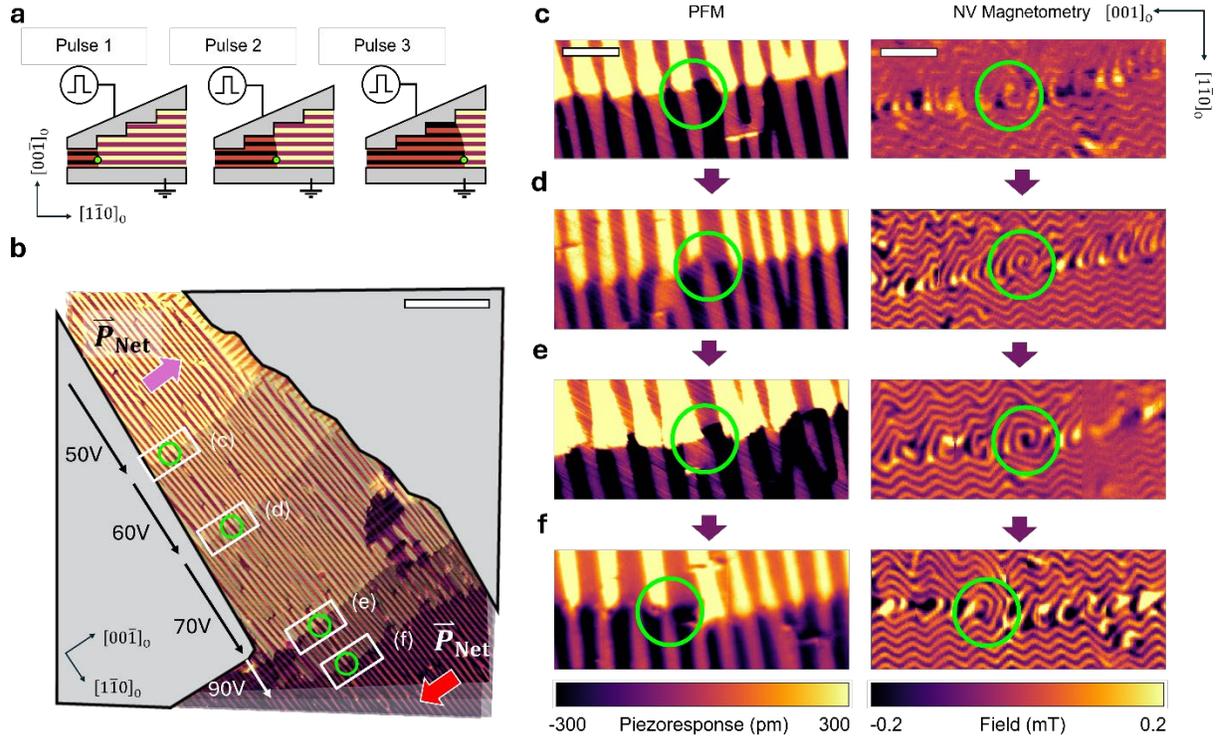

**Figure 2 | Translation of magnetoelectric DWs and topologies down a ratchet racetrack device. a)** A schematic of a magnetoelectric DW ratchet device after consecutive voltage pulses. Grey areas indicate lateral electrodes, and the green circle indicates the location of a magnetoelectric topology at a domain boundary. **b)** An overlay of LPFM images of a ~100 nm BFO film grown on DSO after four consecutive voltage pulses displaying the propagation of magnetoelectric DW. The green circles highlight the location of polar vertex which has been translated down the racetrack. Red and pink arrows designate the net polarization direction $\bar{P}_{Net}$ in different regions of the film. The scalebar is 4 μm. **c-f)** High resolution LPFM and scanning NV magnetometry of regions in (b) marked by white boxes. The green circles note the same vertex shown by the green circles in (b). The scalebar in (c)-(f) is 500 nm. pm, picometers; mT, milliTesla; O, orthorhombic

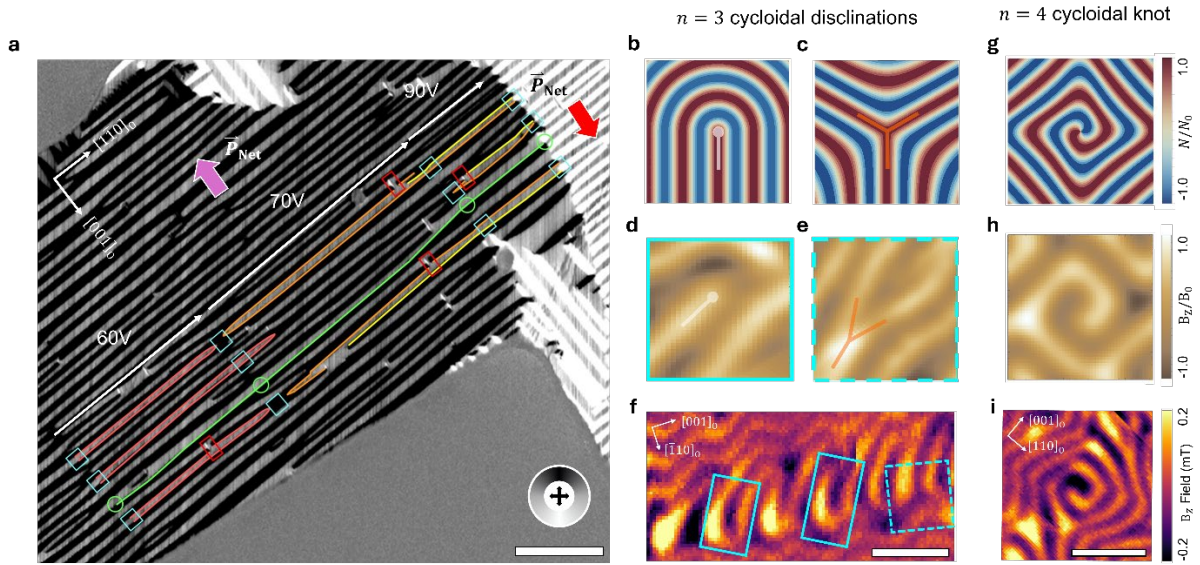

**Figure 3 | Origins and translational stability of magnetoelectric topologies. a)** Greyscale lateral PFM image of the final position of the magnetoelectric DW motion. The net polarization in the virgin and switched regions are noted by red and pink arrows. Trajectories of $n = 3$ and $n = 4$ magnetoelectric topologies are shown by colored lines after sequential electric-field pulses (shown by white arrows). Blue circles indicate the stopping positions of bi-merons, while green circles indicate the stopping positions of vertices. Trajectory line color changes when the topological defect can no longer be unambiguously tracked down the device due to annihilation, merger, or nucleation events. Red boxes indicate defects along the track that do not alter the topology translation. The scalebar in (a) is 2 μm. Néel vector structures of a **b)** $+1/2$ "i" disclination and a **c)** $-1/2$ "y" disclination with their corresponding **d-e)** out-of-plane stray field ($B_Z$) profile for an $n = 3$ cycloidal distillation. Disclinations noted with white and orange lines **f)** Experimentally measured ($B_Z$) of $\pm 1/2$ disclinations along a 180° domain boundary with multiple polar bi-merons. Exemplary $+1/2$ "i" ($-1/2$ "y") disclinations are marked in solid (dashed) boxes. **g)** Computed Néel vector, **h)** computed ($B_Z$), and **i)** experimentally measured ($B_Z$) for an $n = 4$ cycloidal knot topology. The scalebars in (f) and (i) are 250 nm

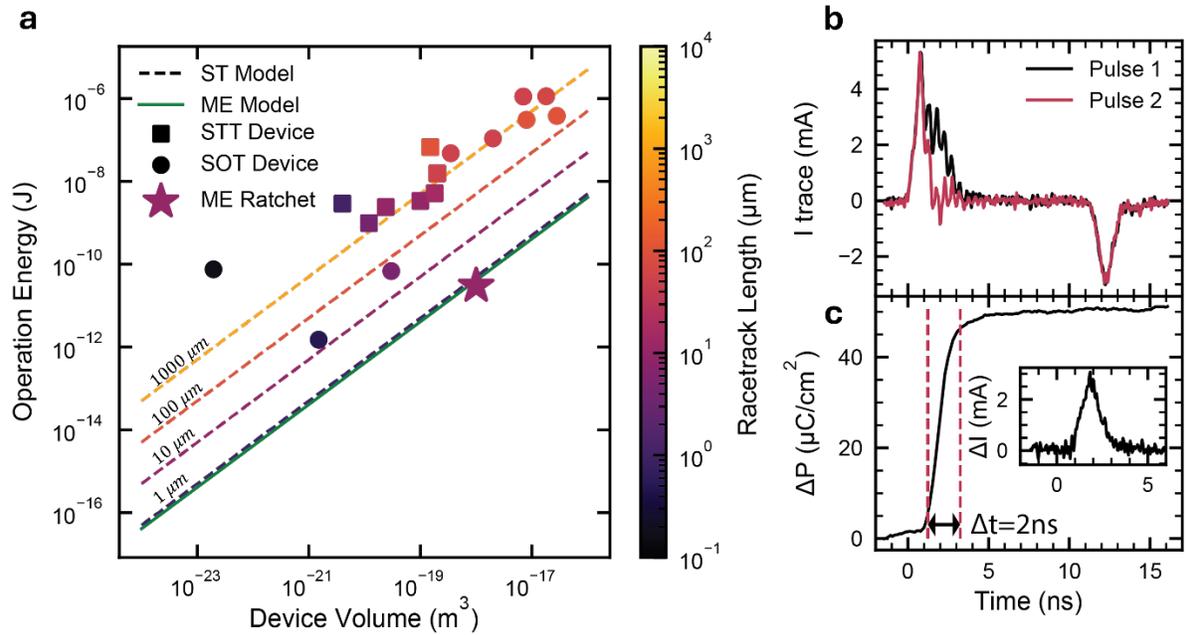

**Figure 4 | Energy consumption and speed comparison. a)** Comparison of the primary operation energy of the electric-field-driven magnetoelectric ratchet device in the current study and exemplary current-driven spin torque systems taken from literature over the last 15 years as a function of device volume. Solid and dashed lines denote magnetoelectric and current-driven spin torque models, respectively. Symbol shape and color refer to DW/spin texture driving mechanisms and racetrack length, respectively. The voltage-driven magnetoelectric ratchet in the current study is plotted with a star. **b)** Dynamic current traces of a single DW traveling down a ratchet DW track, and **c)** the corresponding polarization ($\Delta P$) transient used to extract the magnetoelectric DW and topology velocity. ME, magnetoelectric; ST, spin torque; STT, spin transfer torque; SOT, spin orbit torque.